\documentclass[aps,pre,preprint,groupedaddress,showpacs]{revtex4-1}

\usepackage{amssymb}
\usepackage{amsmath}
\usepackage{graphicx}

\usepackage{enumerate}

\usepackage{graphicx}

\begin{document}

\title{Dynamics and physical interpretation of quasi-stationary states in systems with long-range interactions}


\author{T.~M.~Rocha Filho}
\affiliation{Instituto de F\'\i{}sica and International Center for Condensed Matter Physics\\ Universidade de
Bras\'\i{}lia, CP: 04455, 70919-970 - Bras\'\i{}lia, Brazil}
\author{M.~A.~Amato}
\affiliation{Instituto de F\'\i{}sica and International Center for Condensed Matter Physics\\ Universidade de
Bras\'\i{}lia, CP: 04455, 70919-970 - Bras\'\i{}lia, Brazil}
\author{J.~R.~S.~Moura}
\affiliation{Instituto de F\'\i{}sica\\ Universidade de
Bras\'\i{}lia, CP: 04455, 70919-970 - Bras\'\i{}lia, Brazil}
\author{A.~E.~Santana}
\affiliation{Instituto de F\'\i{}sica and International Center for Condensed Matter Physics\\ Universidade de
Bras\'\i{}lia, CP: 04455, 70919-970 - Bras\'\i{}lia, Brazil}
\author{A.~Figueiredo}
\affiliation{Instituto de F\'\i{}sica and International Center for Condensed Matter Physics\\ Universidade de
Bras\'\i{}lia, CP: 04455, 70919-970 - Bras\'\i{}lia, Brazil}

\begin{abstract}
Although the Vlasov equation is used as a good approximation for a sufficiently large $N$,  Braun and Hepp have showed that
the time evolution of the one particle distribution function of a $N$ particle classical Hamiltonian system with
long range interactions satisfies the Vlasov equation in the limit of infinite $N$.
Here we rederive this result using a different approach allowing a discussion of the role of inter-particle
correlations on the system dynamics.
Otherwise for finite N collisional
corrections must be introduced. This has allowed the a quite comprehensive study of the Quasi Stationary States (QSS)
but many aspects of the physical interpretations of these states remain unclear.
In this paper a proper definition of timescale for long time evolution is discussed and several numerical results are presented,
for different values of $N$.
Previous reports indicates that the lifetimes of the QSS scale as $N^{1.7}$ or even the system
properties scales with $\exp(N)$. However, preliminary results presented here shows indicates that time scale goes as $N^2$ for
a different type of initial condition.
We also discuss how the form of the inter-particle potential determines the convergence of the $N$-particle dynamics to the Vlasov equation.
The results are obtained in the context of following models: the Hamiltonian Mean Field,
the Self Gravitating Ring Model, and a 2-D Systems of Gravitating Particles. We have also provided information of the validity of
the Vlasov equation for finite $N$, i.~e.\ how the dynamics converges to the mean-field (Vlasov) description as $N$ increases and how
inter-particle correlations arise.
\end{abstract}


\maketitle

\section{Introduction}

Long-range interacting systems are characterized by an interaction
potential decaying at long distances as $r^{-\alpha}$ such that $\alpha \leq d$, with
$d$ being the space dimension and may lead to anomalous behavior as
non-Gaussian Quasi-Stationary States (QSS), negative (microcanonical) heat capacity, ensemble inequivalence and
non-ergodicity~\cite{prrev,proc1,proc2,proc3,nos,eplnos}. Examples of systems with long-range interactions include
self-gravitating systems (stars in galaxies and globular clusters), non-neutral plasmas and
two-dimensional vortices~\cite{padma,levin0,levin,levin2,twodflows,fel,buyl,ring1,ring2}. The existence of non-equilibrium non-Gaussian
QSS have been explained by identifying them to stationary states of the Vlasov equation~\cite{liboff}, which describes the dynamics
of the statistical states of such systems.
In a seminal and quite intricate paper~\cite{braun} Braun and Hepp showed that the
time evolution of the one-particle distribution function $f({\bf r},{\bf p},t)$ of a
$N$ particle classical Hamiltonian system with long-range interactions and with a Hamiltonian of the form:
\begin{equation}
H=\sum_{i=1}^N\frac{{\bf p}_i}{2m}+\frac{1}{N}\sum_{i<j=1}^N v({\bf r}_i-{\bf r}_j),
\label{genlongham}
\end{equation}
with ${\bf r}_i$ and ${\bf p}_i$ the position and momentum vectors for particle $i$, respectively,
satisfies the Vlasov equation in the limit $N\rightarrow\infty$:
\begin{equation}
\frac{\partial f}{\partial t}+{\bf p}\cdot\frac{\partial f}{\partial\bf r}+{\bf F}\cdot\frac{\partial f}{\partial\bf p}=0,
\label{vlasoveq}
\end{equation}
where the mean-field force is given by
\begin{eqnarray}
 & & {\bf F}({\bf r},t)=-\frac{\partial}{\partial\bf r}\overline{V}({\bf r},t),\nonumber\\
 & & 
\overline{V}({\bf r},t)=\int v({\bf r}-{\bf r}^\prime) f({\bf r},{\bf p},t) f({\bf r}^\prime,{\bf p}^\prime,t)
\:d{\bf p}\:d{\bf p}^\prime\:d{\bf r}^\prime.
\label{meanfieldfv}
\end{eqnarray}
For finite $N$ Eq.~(\ref{vlasoveq}) is only valid up to terms ${\cal O}(N^{-1})$, and collisional corrections become relevant
(see~\cite{prrev} and references therein for how to obtain kinetic equations for such systems).
The property of vanishing inter-particle correlations in the initial state is consistently propagated by the Hamiltonian dynamics in the
$N\rightarrow\infty$ limit. For practical purposes, the Vlasov equation is used as a good approximation for sufficiently large $N$, its validity
being limited for short times. Although
the lifetime of a QSS have been extensively studied in the literature~\cite{nv1,nv2,nv3,nv4,nv4b,nv5,nv6} many aspects of its physical interpretation and
phenomenology remain unclear. In the present paper we discuss how to properly define a timescale for its long-time evolution which is governed
by collisional corrections to the Vlasov equation leading to the Landau or Balescu-Lenard kinetic equations
(see Ref.~\cite{chavanis0,chavanisa,chavanisb,chavanisc} and references therein).
For homogeneous one-dimensional systems the collision terms of the Landau or Balescu-Lenard equations vanish identically~\cite{chavanis0,sano} and 
higher order corrections must be considered in the kinetic equations, leading to
an evolution timescale of order $N^\delta$ with $\delta>1$ for homogeneous and of order $N$ for inhomogeneous states. For the
Hamiltonian Mean Field (HMF) model (see Section~\ref{convergence} below) a value of $\delta=1.7$ was claimed
by Yamaguchi et al.~\cite{yamaguchi} and by  Bouchet and Dauxois~\cite{bouchet} while Campa et al.~\cite{nv2} and Chavanis~\cite{chavanis2} even reported
a timescale of order $\exp(N)$. This is to be compared to a timescale of order $N^2$ estimated numerically for one-dimensional plasmas~\cite{dawson,rouet}
and to the same scaling for the dynamics of a three level homogeneous initial state for the HMF model presented in Section~\ref{qssvi}.
Also, as we argue below, for most of the QSS a lifetime cannot be defined consistently. In such cases it must be replaced by the notion
of a characteristic time scale (relaxation time).

We also study the scope of validity of the Vlasov equation for finite $N$ and
for some representative models with long-range interactions, i.~e.\ how the dynamics converges to the mean field (Vlasov) description as $N$
increases and how inter-particle correlations arise and how the dynamics depends on them.
The representative models referred are: the Hamiltonian Mean Filed (HMF) model~\cite{hmf},
the self-gravitating ring model~\cite{ring1} and two dimensional self-gravitating particles~\cite{2dgrav}.
There are a few approaches to deduce the Vlasov equation from first principles for long-range forces~\cite{braun,chavanis1,prrev}
but we chose to use the one by Balescu which permits to explicitly estimate the order of magnitude of many contributions,
and most importantly that of inter-particle correlations~\cite{balescu1,balescu2}. For sake of completeness and for the present paper
to be self-contained, we present this approach in the appendices. The Vlasov equation is obtained from a resummation of different contributions of
a diagrammatic expansion. It is shown that provided all inter-particle correlations are dynamically created, they do not alter
the validity of the Vlasov equation. This means that deviations from the solution of the Vlasov equation are not due to
the building up of the correlations with time, since their order of magnitude is preserved by the dynamics, but to the secular accumulation
of small collisional corrections of order $1/N$ (see Section~\ref{secrole}). The one-particle distribution function at any point of the evolution
can thus be used as the initial condition to solve the Vlasov equation, which is valid for a given time span starting at this value of time.
This is important for the interpretation of the QSSs and their characteristic lifetimes.

This paper is structured in the following way: Section~\ref{vlasovdeduc} presents the deduction of the Vlasov equation from Balescu's dynamics of correlations
approach. In Section~\ref{convergence} we discuss the effect of the explicit form of the inter-particle potential on the convergence of the $N$-particle dynamics
to the solution of the Vlasov equation. Section~\ref{secrole} discusses
the behavior of inter-particle correlation with time from simulation data and Section~\ref{qssvi} the physical
interpretation of quasi-stationary states in the light of previous Sections. We close the paper with some concluding remarks in Section~\ref{conclu}.
Appendix~\ref{appendixa} presents the dynamics of correlations formalism of Balescu and Appendix~\ref{appendixb} how it can be extended for
self-gravitating systems with different masses.

\section{Dynamics of correlations and the Vlasov equation}
\label{vlasovdeduc}

The diagrammatic approach for the solution of the $N$-particle Liouville equation was used by Balescu to deduce the Vlasov equation for plasmas as a lower order
evolution equation in the plasma parameter\cite{balescu1,balescu2}. In this Section we show that the Kac factor $1/N$ in the interaction potential leads
naturally to the Vlasov equation by allowing to select the diagrams contributing to the solution of the Liouville equation.
The long-range nature of the interaction manifests itself in the non-vanishing of the mean-field potential in the $N\rightarrow\infty$ limit
(see the discussion in Section 5.3 of Ref.~\cite{balescu1}). We believe that the present approach is more physically appealing and allow to assess explicitly
the different relevant contributions to the Vlasov equation.

We consider here a system of identical particles with Hamiltonian with Cartesian coordinates:
\begin{equation}
H=\sum_{i=1}^{N}\frac{p_{i}^{2}}{2m_i}+\frac{1}{N}\sum_{i<j=1}^{N}V_{ij},
\label{hamiltonian}
\end{equation}%
where $V_{ij}\equiv V(|\mathbf{r}_{i}-\mathbf{r}_{j}|)$ the pair interaction potential
for particles $i$ and $j$, and ${\bf p}_i=m_i{\bf v}_i$, $\mathbf{r}_{i}$, $m_i$ the momentum, position and mass of particle $i$,
respectively. For identical particles, and in order to simplify the presentation, we chose the unit of mass such that $m_i=1$.

The probability that particle $i$ is in the phase volume $d{\bf r}_i d{\bf v}_i$ is written as
(we use the notation of Ref.~\cite{liboff}):
\begin{equation}
f_{N}(1,2,\ldots ,N)\:d1\:d2\cdots dN,
\label{probfn}
\end{equation}%
where $f_{N}$ is the $N$-particle distribution function, and we write $1$
for $\mathbf{r}_{1},\mathbf{v}_{1}$, $d1\equiv d\mathbf{r}_{1}d\mathbf{v}_{1}$, $2$
for $\mathbf{r}_{2},\mathbf{v}_{2}$, $d2\equiv d\mathbf{r}_{2}d\mathbf{v}_{2}$, and so on.
We impose that $f_N$ is a symmetric function with respect to particle interchange.The $s$-particle
distribution function is defined by
\begin{equation}
f_s(1,\ldots,s)=\int f_N(1,\ldots,N)\:d(s+1)\cdots dN.
\label{fsdef}
\end{equation}
For a closed system $f_{N}$ satisfies the Liouville equation:
\begin{equation}
\frac{\partial f_{N}}{\partial t}=\hat L_{N}f_{N},
\label{liouvilleeq}
\end{equation}
with the Liouville operator given by
\begin{eqnarray}
\hat L_{N} &\equiv &\left\{ H,\hspace{8pt}\right\} =\sum_{i=1}^{N}
\left\{ \frac{\partial H}{\partial \mathbf{r}_{i}}\cdot \frac{\partial }{\partial \mathbf{v}_{i}}
-\frac{\partial H}{\partial \mathbf{v}_{i}}\cdot \frac{\partial }{\partial \mathbf{r}_{i}}\right\}
\nonumber \\
&=&\hat L_{0}+\delta\hat L,
\label{delldl}
\end{eqnarray}
where
\begin{eqnarray}
\hat L_{0} & = & \sum_{i=1}^{s}\mathbf{v}_i\cdot \frac{\partial }{\partial \mathbf{r}_i},
\nonumber\\
\delta\hat L & = & \frac{1}{N}\sum_{i\neq j=1}^{s}\frac{\partial V_{ij}}{\partial{\bf r}_i}\cdot
\left(\frac{\partial}{\partial\mathbf{v}_i}-\frac{\partial }{\partial \mathbf{v}_j}\right).
\end{eqnarray}
A formal expression for the solution of the Liouville equation is given by~\cite{balescu1}:

\begin{equation}
f_{N}(t)=-\frac{1}{2\pi i}\oint
dz\:e^{-izt}(\hat L-z)^{-1}f_{N}(0)=e^{-iLt}f_{N}(0),\hspace{4mm}t>0.
\label{laplsol}
\end{equation}
Using iteratively the identity
\begin{equation}
(\hat L-z)^{-1}\equiv(\hat L_0-z)^{-1}\left[1-\delta\hat L(\hat L-z)^{-1}\right],
\label{identresolv}
\end{equation}
results in a series (perturbative) solution of the Liouville equation:
\begin{equation}
|f_N(t)\rangle=-\frac{1}{2\pi i}\oint dz\:e^{-izt}\sum_{n=0}^{\infty}
(\hat L_{0}-z)^{-1}\left[- \delta\hat L(\hat L_{0}-z)^{-1}\right]^n|f_N(0)\rangle,
\label{rho}
\end{equation}
where we introduced a Bra-Ket notation here and in what follows (see Ref.~\cite{liboff}).

\begin{figure}[ht]
\begin{center}
\scalebox{0.6}{{\includegraphics{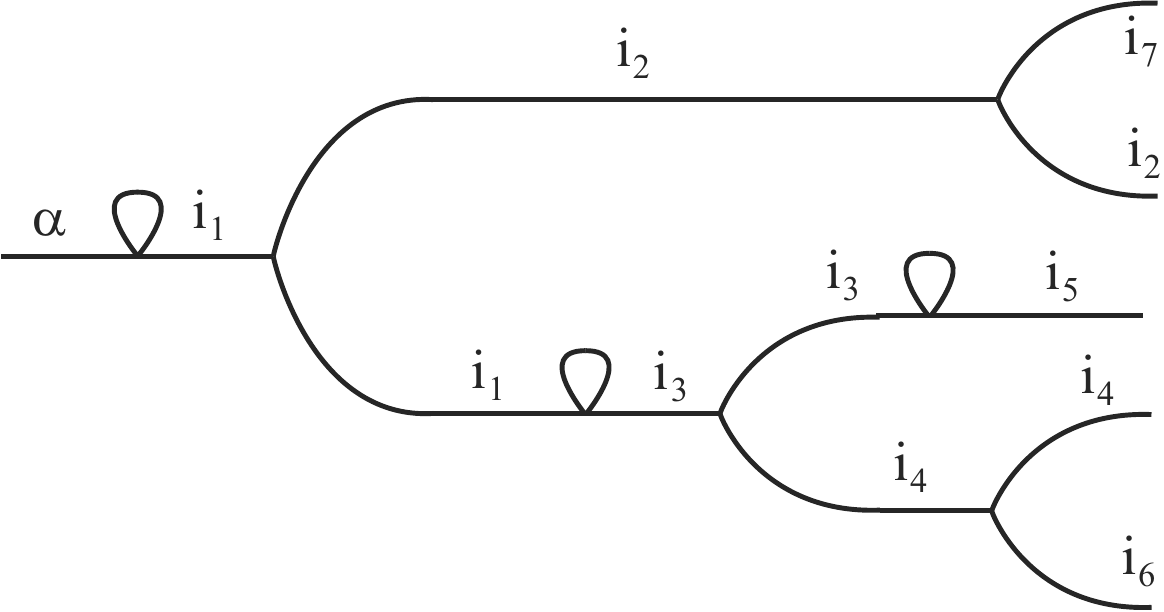}}}
\end{center}
\caption{Example of diagram occurring in Eq.~(\ref{braketexp}) contributing to the Fourier coefficient $a_{\bf k}$.}
\label{exvlas}
\end{figure}

Let us consider the spatial Fourier transform of $f_N$ in $d$ spatial dimensions, and the corresponding inverse transform:
\begin{eqnarray}
 & & a_{\{\bf k\}}\equiv a_{{\bf k}_1,\ldots,{\bf k}_N}({\bf v},t)=\int f_{N}({\bf r},{\bf v},t)\: \exp\left(\sum_{j=1}^N{\bf k}_j\cdot{\bf r}_j\right)\: d^N{\bf r},
\nonumber\\
 & & f_{N}({\bf r},{\bf v},t)=\frac{1}{(2\pi)^{dN}}\int a_{\{\bf k\}}\exp\left(-\sum_{j=1}^N{\bf k}_j\cdot{\bf r}_j\right)\: d^N{\bf v},
\label{fourexp}
\end{eqnarray}
or equivalently
\begin{equation}
|f_N\rangle=\sum_{\{{\bf k}\}}a_{\{{\bf k}\}}({\bf v},t)|\{{\bf k}\}\rangle,\hspace{10mm}
a_{\{{\bf k}\}}({\bf v},t)=\langle\{{\bf k}\}|f_N\rangle.
\label{fourexpbraket}
\end{equation}
For some model systems, as the ring or the HMF models~\cite{ring1,ring2,hmf}, the configuration space is periodic, and the Fourier transform
is replaced by a Fourier series. By integrating Eq.~(\ref{fourexp}) over particles $2$ through $N$ the only remaining contributions
are those with at most one non-vanishing wave-vector:
\begin{equation}
f_1({\bf r},{\bf v})=a_0({\bf v},t)+\int a_{\bf k}({\bf v},t) e^{-i{\bf k}\cdot{\bf r}}\: d{\bf k},
\label{r3}
\end{equation}
where we used the identity
\begin{equation}
\int e^{i{\bf k}\cdot{\bf r}}d{\bf r}=(2\pi)^d\delta({\bf k}),
\label{r2}
\end{equation}
with $\delta({\bf k})$ the $d$-dimensional Dirac delta function. Therefore only the time evolution of coefficients with one or none non-vanishing wave-vectors
must be determined in order to obtain a kinetic equation.
We explicitly
obtain all diagrams contributing to the leading order in $1/N$ as follows. First any possible diagram contributing to $a_0$ must start with a  vertex of type C
(see Fig.~\ref{figA1} in Appendix~\ref{appendixa}) and it would necessarily act on the right on a Fourier coefficient describing a correlation,
which is at most of order $N^{-1}$. Since we already have a $N^{-1}$ factor from the vertex,
for such a diagram to be independent of $N$ we should sum over two particles (each sum contributing a factor $N$),
but this would lead to vanishing surface terms. Therefore the diagram is at most of order $1/N$.
Following this reasoning, adding more vertices to the diagram cannot result in a diagram independent of $N$.
As a consequence $a_0$ is constant in time up to this order.
Similarly, the only non-vanishing contributions to $a_{\bf k}$ are diagrams formed by any combination of vertices of
type D and F (Fig.~\ref{exvlas} shows an example of such diagrams).

All diagrams contributing to $a_{\bf k}$ are put into two classes: those beginning at left with vertex F (class F) and those beginning with a vertex D (class D).
Removing the first vertex F on the left in class F results in diagrams contributing to $a_{\bf k}({\bf v}_j,t)$, and similarly
removing the first vertex D on the left in class D gives those contributions to $a_{\bf k^\prime}({\bf v}_\alpha,t)a_{{\bf k}-{\bf k}^\prime}({\bf v}_j,t)$.
Thence the formal solution has the following general structure:
\begin{eqnarray}
 & & a_{\bf k}({\bf v},t)=-\frac{1}{2\pi i}\oint\: e^{-izt}\:\frac{1}{i\left({\bf k}\cdot{\bf v}_\alpha-z\right)}
 \left[a_{\bf k}(0)-\frac{i}{N}\sum_j V_{\bf k}\:{\bf k}\cdot\partial_{\alpha j}\:{\cal F}_{\bf k}\right.
\nonumber\\
 & & \hspace{10mm}\left.-\frac{i}{N}\sum_j \int V_{|{\bf k}-{\bf k}^\prime|}({\bf k}-{\bf k}^\prime)\partial_{\alpha j}
{\cal D}_{{\bf k}^\prime,{\bf k}-{\bf k}^\prime}\: d{\bf k}^\prime \right]\:dz,
\label{solgenstr}
\end{eqnarray}
where ${\cal D}_{{\bf k}^\prime,{\bf k}-{\bf k}^\prime}$ and ${\cal F}_{\bf k}$ represent the contributions from
class D and F above after removing the first vertex at its left in each diagram.
Differentiating both sides with respect to time introduces a factor $-iz=i({\bf k}\cdot{\bf v}_\alpha-z)-i{\bf k}\cdot{\bf v}_\alpha$ in the integrand, and yields
after some manipulations:
\begin{eqnarray}
 \lefteqn{\frac{\partial}{\partial t} a_{\bf k}({\bf v},t) =-i{\bf k}\cdot{\bf v}_\alpha\:a_{\bf k}({\bf v}_\alpha,t)
+V_{\bf k}\:i{\bf k}\cdot\frac{\partial}{\partial{\bf v}_\alpha}\int a_{\bf k}({\bf v}^\prime,t)\:d{\bf v}^\prime}
\nonumber\\
 & & +\frac{\partial}{\partial{\bf v}_\alpha}\cdot\int d{\bf k}\int d{\bf v}^\prime\:iV_{|{\bf k}-{\bf k}^\prime|}({\bf k}-{\bf k}^\prime)
a_{{\bf k}^\prime}({\bf v}_\alpha,t)a_{{\bf k}-{\bf k}^\prime}({\bf v}^\prime,t).
\label{dersolgen}
\end{eqnarray}
Using Eq.~(\ref{r3}) and the results in Appendix~\ref{appendixa} we finally obtain:
\begin{eqnarray}
\partial _{t}f(\mathbf{r}_{\alpha },\mathbf{v}_{\alpha };t)&=&-\mathbf{v}%
_{\alpha }\cdot \nabla f(\mathbf{r}_{\alpha
},\mathbf{v}_{\alpha};t) + \partial _{\alpha
}f(\mathbf{r}_{\alpha },\mathbf{v}_{\alpha };t)\cdot
\frac{\partial}{\partial{\bf v}_\alpha}\nonumber \\
&&\times \int d\mathbf{r}^\prime d\mathbf{v}^\prime\: V(|\mathbf{r}_{\alpha }-\mathbf{r}^\prime|)
f(\mathbf{r}_{\alpha },\mathbf{v}_{\alpha };t).
\label{vlasovfinal}
\end{eqnarray}
Equation~(\ref{vlasovfinal}) is the desired form for a Vlasov equation. It can also be obtained in the same way for
a system of non-identical self-gravitating particles as explained in Appendix~\ref{appendixb}. We now discuss in the next Section how to
consider the contributions of correlation (possibly) present in the initial state.


\subsection{Order of magnitude of the correlations in $N$.}

\begin{figure}[ht]
\begin{center}
\scalebox{1.0}{{\includegraphics{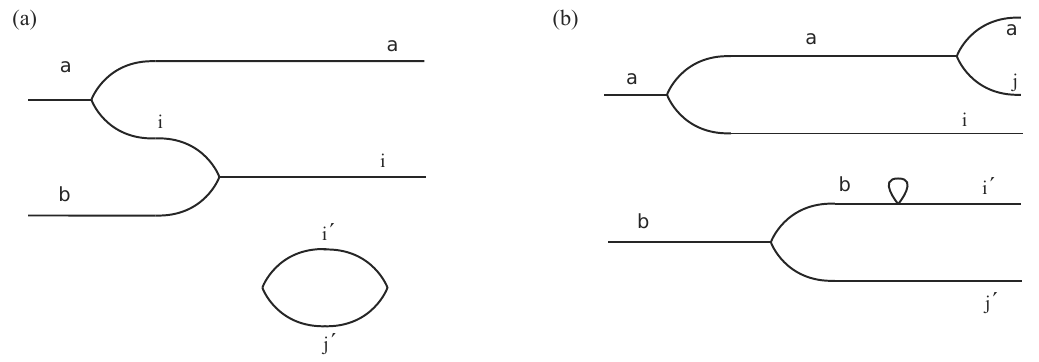}}}
\end{center}
\caption{Examples of diagram contributing to the coefficient $a_{{\bf k}_\alpha,{\bf k}_\beta}$.}
\label{factex}
\end{figure}

Due to the linearity of the Liouville equation Eq.~(\ref{liouvilleeq}) as well of the Fourier coefficients $a_ {\{\bf k\}}({\bf v},t)$,
the order of magnitude of correlations for the initial distributions are kept constant by the time evolution. Therefore, if all correlations in the system are dynamically created,
they are expect to have the same dependency on $N$ as in the final state, i.~e.\ the thermodynamical equilibrium. It was shown in Ref.~\cite{chavanis1} that the microcanonical
equilibrium distribution factorizes up to terms of order $N^{-1}$:
\begin{equation}
f_N^{\rm eq}(1,\ldots,N)=\prod_{i=1}^N f_1^{\rm eq}({\bf r}_i,{\bf v}_i)+{\cal O}\left(\frac{1}{N}\right).
\label{facteq}
\end{equation}
The Fourier transform of the factorized equilibrium distribution function leads to
\begin{equation}
a^{\rm eq}_{\{\bf k\}}({\bf v})
\label{factcoeff}\equiv a^{\rm eq}_{{\bf k}_1,\ldots,{\bf k}_N}({\bf v}_1,\ldots,{\bf v}_N)=\prod_{i=1}^N a^{\rm eq}_{1,{\bf k_i}}({\bf v}_i)+{\cal O}\left(\frac{1}{N}\right).
\end{equation}
Therefore we assume that the correlation pattern $a_{[{\bf k}_1,\ldots,{\bf k}_l]}$ in Eq.~(\ref{jan29.1}) is at least one order of magnitude less than the purely
factored term $a_{{\bf k}_1}a_{{\bf k}_2}\cdots a_{{\bf k}_l}$.
Figure~\ref{factex} shows two examples of different contributions
to the coefficient $a_{{\bf k}_\alpha,{\bf k}_\beta}$ with diagrams composed of four vertices of type A, B, C and D contributing with a factor $N^{-4}$.
From property (iv) in Appendix~\ref{apa2}, we see that in Figure~\ref{factex}a either $j'=\alpha$ or $j'=\beta$ in order to yield a non-vanishing contribution. Therefore we have three summations
over particle indices: $i$, $j$ and $i'$, contributing with a factor $N^3$. The diagram is thus of order $N^{-1}$. A similar analysis for the diagram in Fig.~\ref{factex}b
implies that it is independent of $N$ (four vertices and four particle indices). Therefore in Fig.~\ref{factex} diagram (a) is negligible while (b) must be retained. Diagram
(a) creates a correlation among particles $\alpha$ and $\beta$ while diagram $(b)$ does not.
From these examples it is straightforward to see that only a succession of vertices of type D and F contributes to any reduced distribution
function $f_s$, and therefore cannot create any correlations among particles $1$ through $s$. This proves that the factorization of the distribution
functions is maintained by the dynamics up to terms of order $N^{-1}$.
More importantly, this result implies that the one-particle distribution function at any stage of the time evolution can be used as the initial condition
for the Vlasov equation, which then from this new starting point slowly deviates from the finite $N$ dynamics due to the cumulative secular effects of
lower order collisional terms.

\section{Convergence to the Vlasov limit for some representative models}
\label{convergence}

In order to show how different interaction potentials lead to different convergence speed to the Vlasov (mean-field) dynamics
we consider three different one-dimensional models with long-range interactions extensively studied in the literature:
the Hamiltonian Mean Field Model (HMF), with Hamiltonian~\cite{hmf1}:
\begin{equation}
H=\frac{1}{2}
\sum_{i=1}^N p_{i}^{2}+\frac{1}{2N}\sum_{i,j=1}^N
[1-\cos(\theta _{i}-\theta _{j})],
\label{HMF}
\end{equation}
where $\theta _{i}$ is the position angle of particle $i$ and $p_{i}$ its conjugate momentum,
the Ring model~\cite{ring1}:
\begin{equation}
H=\frac{1}{2}\sum_{i=1}^N p_i^2-\frac{1}{2N}\sum_{i,j=1}^N\frac{1}{\sqrt{2}\sqrt{1-\cos(\theta_i-\theta_j)+\epsilon}},
\label{RMham}
\end{equation}
where $\epsilon$ is a softening parameter introduced to avoid the zero distance divergence in the pair interaction potential,
and the infinite sheet model in three dimensions, describing $N$ infinite planes with constant mass density with motion only
along the $x$ axis~\cite{sheet}:
\begin{equation}
H=\frac{1}{2}\sum_{i=1}^N p_i^2+\frac{1}{2N}\sum_{i,j=1}^N \left|x_i-x_j  \right|,
\label{sheetham}
\end{equation}
where $x_i$ is $i$-th particle coordinate.
We also consider a two-dimensional system  with $N$ identical particles with unit mass and unit gravitational constant
with Hamiltonian~\cite{levin0}:
\begin{equation}
H=\sum_{i=1}^N \frac{{\bf p}_i^2}{2}+\frac{1}{2N}\sum_{i,j=1}^N\log\left({\left|{\bf r}_{ij}\right|+\epsilon}\right)\:\:,
\label{selfgravham}
\end{equation}
where ${\bf r}_{ij}$ is the vector from particle $i$ to particle $j$ and $\epsilon$ a softening parameter.
The so-called Kac factor $N^{-1}$ is again introduced with a change of time units (see Ref.~\cite{kiessling} for a discussion of its interpretation and formal
results valid for self-gravitating systems).

\begin{figure}[ptb]
\begin{center}
\scalebox{0.3}{\includegraphics{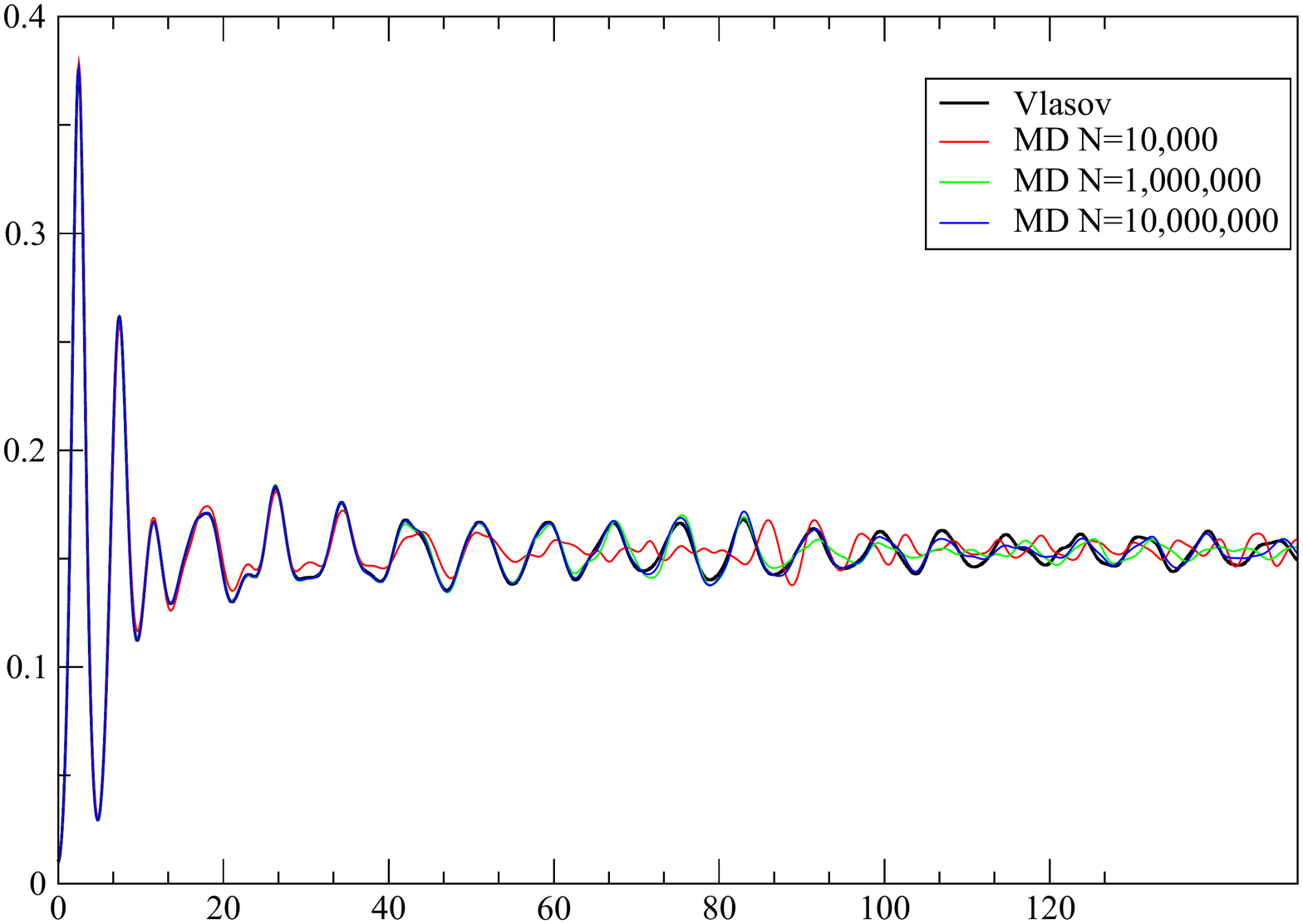}}
\end{center}
\caption{Kinetic energy for Vlasov and MD simulations for the HMF model.
Initial conditions as given in Eq.~(\ref{wbinitheta}) with $p_0=0.25$ and $\theta_0=4.0$.
Time steps are $\Delta t=0.01$ for Vlasov simulation and $\Delta t=0.1$ for MD simulations.}
\label{hmvconv}
\end{figure}

\begin{figure}[ptb]
\begin{center}
\scalebox{0.4}{\includegraphics{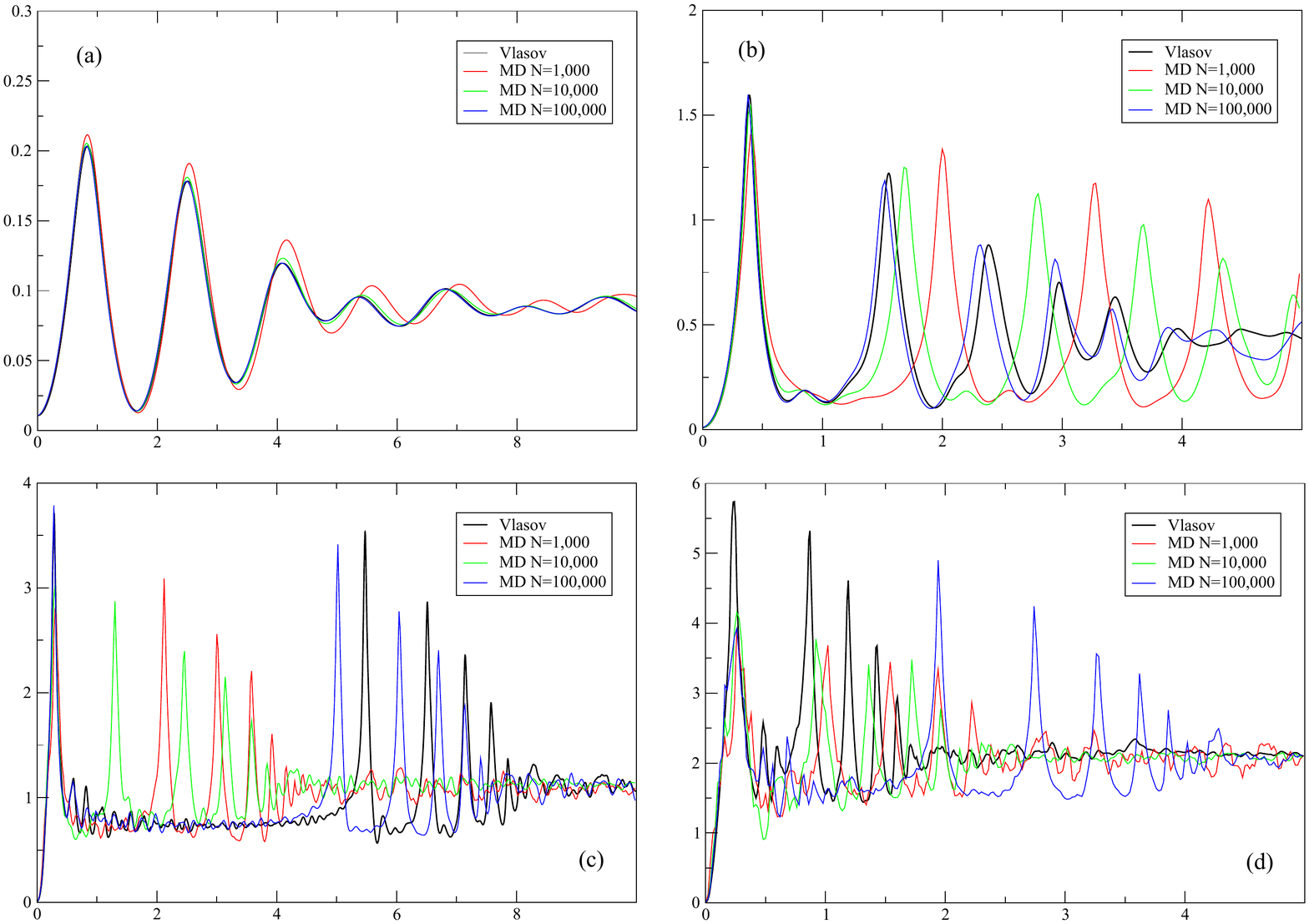}}
\end{center}
\caption{Kinetic energy for Vlasov and MD simulations for the ring model with $\epsilon=10^{-1}$ (panel a),
$\epsilon=10^{-2}$ (panel b), $\epsilon=10^{-3}$ (panel c), $\epsilon=10^{-4}$ (panel d).
The initial condition is a waterbag state with $p_0=0.25$ and $\theta_0=1.0$. Time steps for Vlasov and MD simulations
are $\Delta t=10^{-3}$.}
\label{ringconv}
\end{figure}

\begin{figure}[ptb]
\begin{center}
\scalebox{0.3}{\rotatebox{0}{\includegraphics{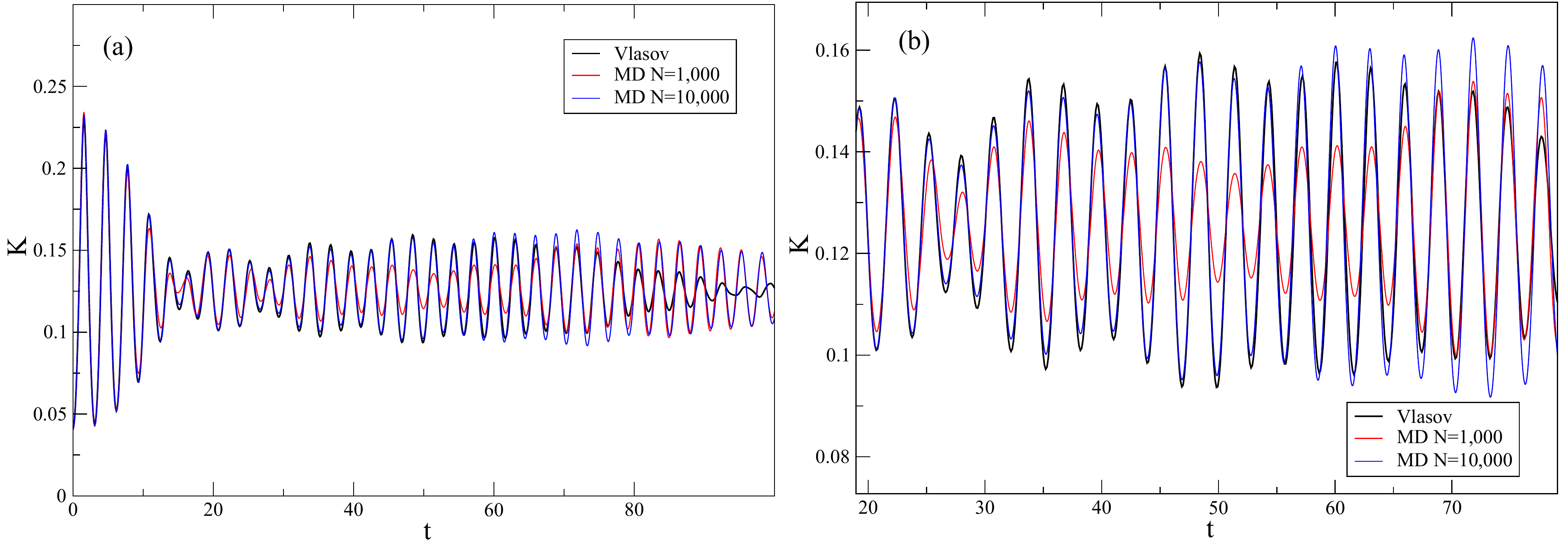}}}
\end{center}
\caption{Panel (a): Kinetic energy for Vlasov and MD simulations for the self-gravitating sheets model. Panel (b) shows the region where
the MD simulation deviates significantly from the Vlasov solution.}
\label{sheetconv}
\end{figure}

All simulations for the one-dimensional models where performed starting from a waterbag initial condition defined by
\begin{equation}
f_0(p,\theta)=\left\{
\begin{array}{ll}
{1}/{2p_0\hspace{1pt}\theta_0}, & \mbox{if $-p_0<p<p_0$ and $0<\theta<\theta_0$,}\\
0, & \mbox{otherwise},
\end{array}
\right.
\label{wbinitheta}
\end{equation}
for the HMF and ring models, and
\begin{equation}
f_0(p,x)=\left\{
\begin{array}{ll}
{1}/{4p_0\hspace{1pt}x_0}, & \mbox{if $-p_0<p<p_0$ and $-x_0<\theta<x_0$,}\\
0, & \mbox{otherwise},
\end{array}
\right.
\label{wbinix}
\end{equation}
for the sheets model. Values chosen for $p_0$, $\theta_0$ and $x_0$ are indicated in the respective figure captions.
For the two-dimensional self-gravitating systems the initial distribution is given by all particles at rest
and homogeneously distributed in a circular shell of inner and outer radius $R_1$ and $R_2$ respectively.
Molecular Dynamics simulations (MD) were performed using a fourth-order simplectic integrator~\cite{yoshida,eu2} for the HMF and ring models,
and an event driven algorithm for the sheets model~\cite{rapaport}. Time steps for simplectic integration are also indicated
in the figure captions. All Vlasov simulations are performed using the approach described in Ref.~\cite{vlasoveu}, with a numerical grid
with $4096\times4096$ points in the $p,\theta$ or $p,x$ one-particle phase space. The form of the pair interaction potential for the HMF
model is such that the simulation time for MD simulations scales with the number of particles $N$ and thence simulations with a great number of particles
are feasible. Figure~\ref{hmvconv} shows the kinetic energy for the HMF model obtained from the solution of the Vlasov equation
and MD simulations with some different values of $N$, with a very good agreement already for $N=10,000$ up to $t\approx20.0$. The time interval
for which finite $N$ and Vlasov solutions agree increases with $N$, as expected. Simulation for the ring model are shown in figure~\ref{ringconv}
for some values of the softening parameter $\epsilon$, and convergence toward the Vlasov values of the Kinetic energy gets clearly
slower for smaller $\epsilon$, as the interaction gets stronger at short distances where collisional effects are more important.
For the self-gravitating sheets model, an increasing agreement with increasing $N$ is also obtained as given in figure~\ref{sheetconv}.

The convergence to a mean-field description is thus strongly affected by the short range part of the force. The stronger the latter the more important
are the collisional effects, and the smaller the agreement time window of the Vlasov equation with the finite $N$ dynamics.

\section{Evolution of inter-particle correlations with time}
\label{secrole}

\begin{figure}[ptb]
\begin{center}
\scalebox{0.3}{\rotatebox{0}{\includegraphics{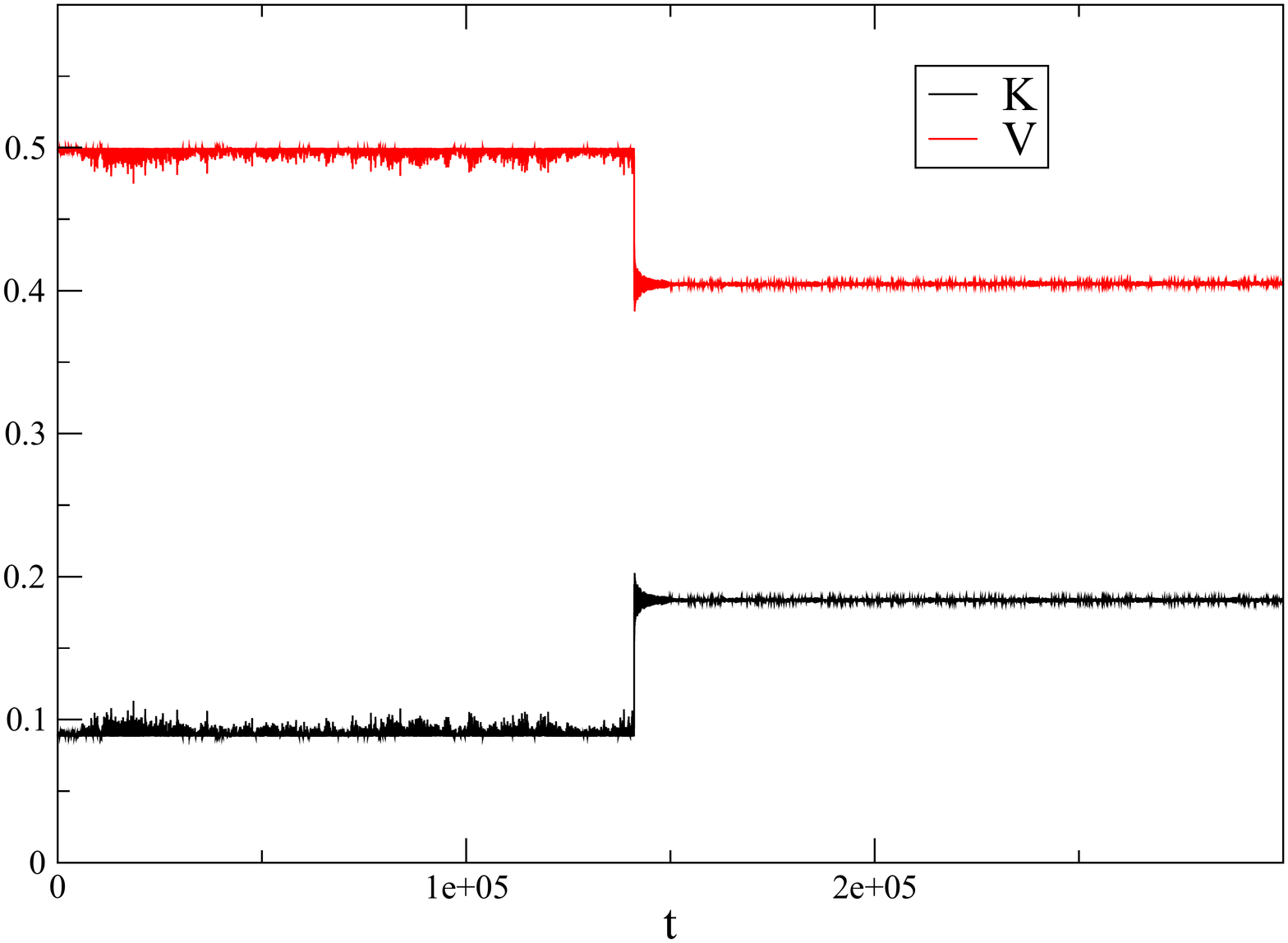}}}
\end{center}
\caption{Kinetic $K$ and potential $V$ energies per particles for the HMF model with $N=262,144$ particles waterbag initial condition with $p_0=0.726$
and $\theta_0=2\pi$, corresponding to zero initial magnetization and total energy per particle $E=0.5879$.}
\label{energies_kurts}
\end{figure}

\begin{figure}[ptb]
\begin{center}
\scalebox{0.35}{\rotatebox{0}{\includegraphics{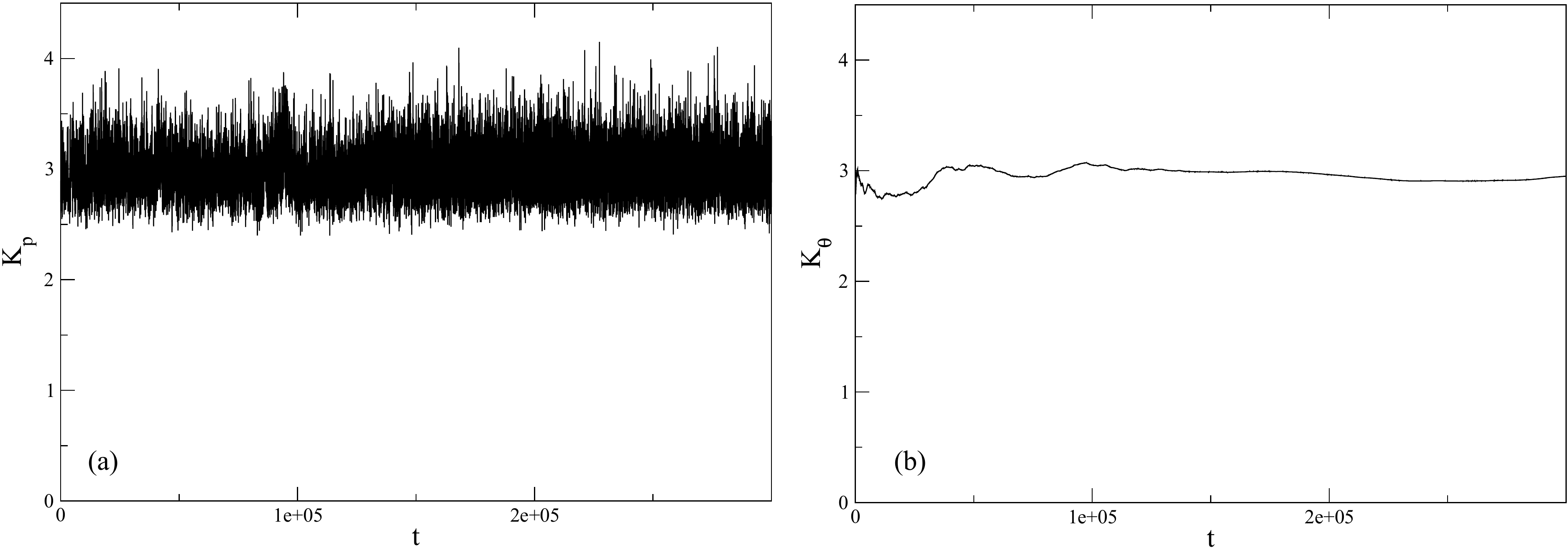}}}
\end{center}
\caption{Kurtosis for the variables in Eq.~(\ref{reducedvars}) corresponding to the same simulation as in figure~\ref{energies_kurts}.
Panel (a): Kurtosis for the sum of momenta variables. Panel (b): Kurtosis for the sum of position variables.}
\label{kurts}
\end{figure}

\begin{figure}[ptb]
\begin{center}
\scalebox{0.3}{\rotatebox{0}{\includegraphics{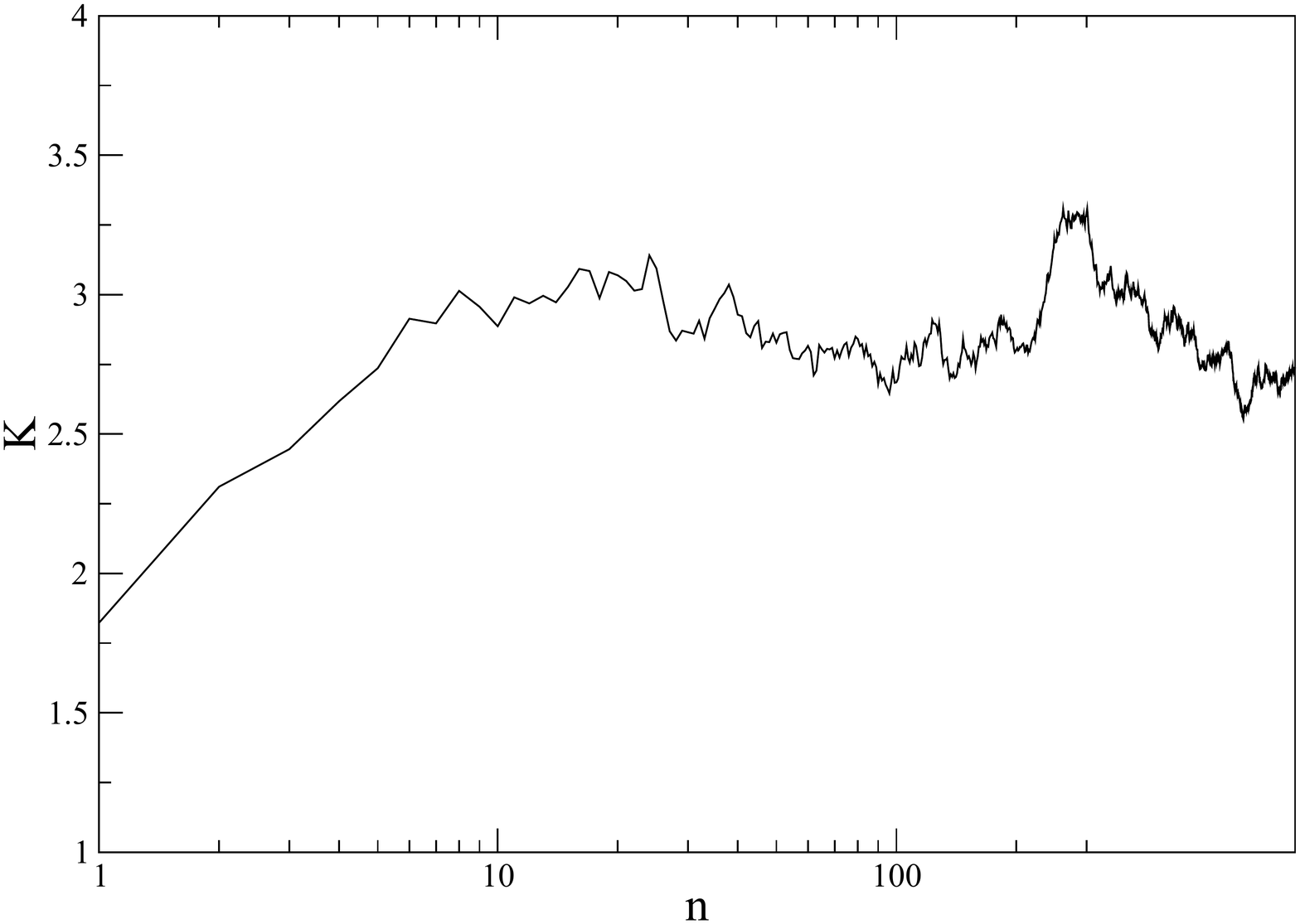}}}
\end{center}
\caption{Kurtosis for a sum of random number variables.}
\label{summedvars2}
\end{figure}

Even though the Vlasov equation can be deduced from the BBGKY hierarchy by imposing uncorrelated particles,
i.~e.\ a fully factored $N$-particle distribution function,
the results in Section~\ref{vlasovdeduc}  stresses the fact that the order of magnitude of the correlations are
preserved by the dynamics and that provided all inter-particle correlations are dynamically created (see also Ref.~\cite{balescu2})
the Vlasov equation is valid at
any stage of the evolution by considering the one particle distribution function at this given time as the initial condition for the Vlasov equation.
From this point on its solution will secularly deviate from the distribution function for finite $N$ due to
collisional terms of order $1/N$ or lower (see Section~\ref{qssvi}).
In order to illustrate this point, let us consider the HMF model with a homogeneous (waterbag) initial condition.
A measure of inter-particle correlations is obtained by partitioning the set of all $N$ particles into $N/M$ groups with $M$ particles each
(we suppose $N$ is divisible by $M$). Then we define the variables:
\begin{eqnarray}
y_k & = & \sum_{i=1}^{M} \theta_{(k-1)m+i},\nonumber\\
z_k & = & \sum_{i=1}^{M} p_{(k-1)m+i},
\label{summedvars}
\end{eqnarray}
and the reduced variables
\begin{eqnarray}
\tilde{y}_k & = & \left(y_k-\langle y\rangle\right)/\sigma_y,\nonumber\\
\tilde{z}_k & = & \left(z_k-\langle z\rangle\right)/\sigma_z,
\label{reducedvars}
\end{eqnarray}
where $\langle\cdots\rangle$ stands for the statistical average and $\sigma_y$ and $\sigma_z$ are the standard deviations of $y$ and $z$, respectively.
If particles are uncorrelated the central limit theorem states that the distribution of $y_k$ and $z_k$ tend
to a Gaussian. The Kurtosis of a distribution if defined as the fourth moment of the reduced variables in Eq.~(\ref{reducedvars}) and
is exactly $3$ for any Gaussian distribution. Figure~\ref{energies_kurts} shows the kinetic and potential energies per particle for a simulation
with $N=262,144$ particles with a homogeneous initial
waterbag state with total energy per particle $E=0.5879$ such that the initial state is stable but close to energy value where the
waterbag state becomes unstable (see Section~\ref{qssvi} below).
The crossover from the homogeneous waterbag state to the final thermodynamical equilibrium is clearly visible in the figure.
Figure~\ref{kurts} shows the Kurtosis ${\cal K}_p$ and ${\cal K}_\theta$ for the variables $\tilde y$ and $\tilde z$ with $M=512$
along the time evolution including the destabilization of the waterbag state and the subsequent evolution towards the final thermodynamical equilibrium.
No substantial deviation from the value corresponding to uncorrelated particles is observed.
This must be compared to the Kurtosis of 1000 realizations of the sum of $n$ random numbers
obtained from a good random number generator~\cite{knuth} as a function of $n$ and shown in figure~\ref{summedvars2}.
This shows that inter-particle correlations are in fact small all along the time evolution, with an order of magnitude preserved by the dynamics, as expected.

\begin{figure}[ptb]
\begin{center}
\scalebox{0.6}{\rotatebox{0}{\includegraphics{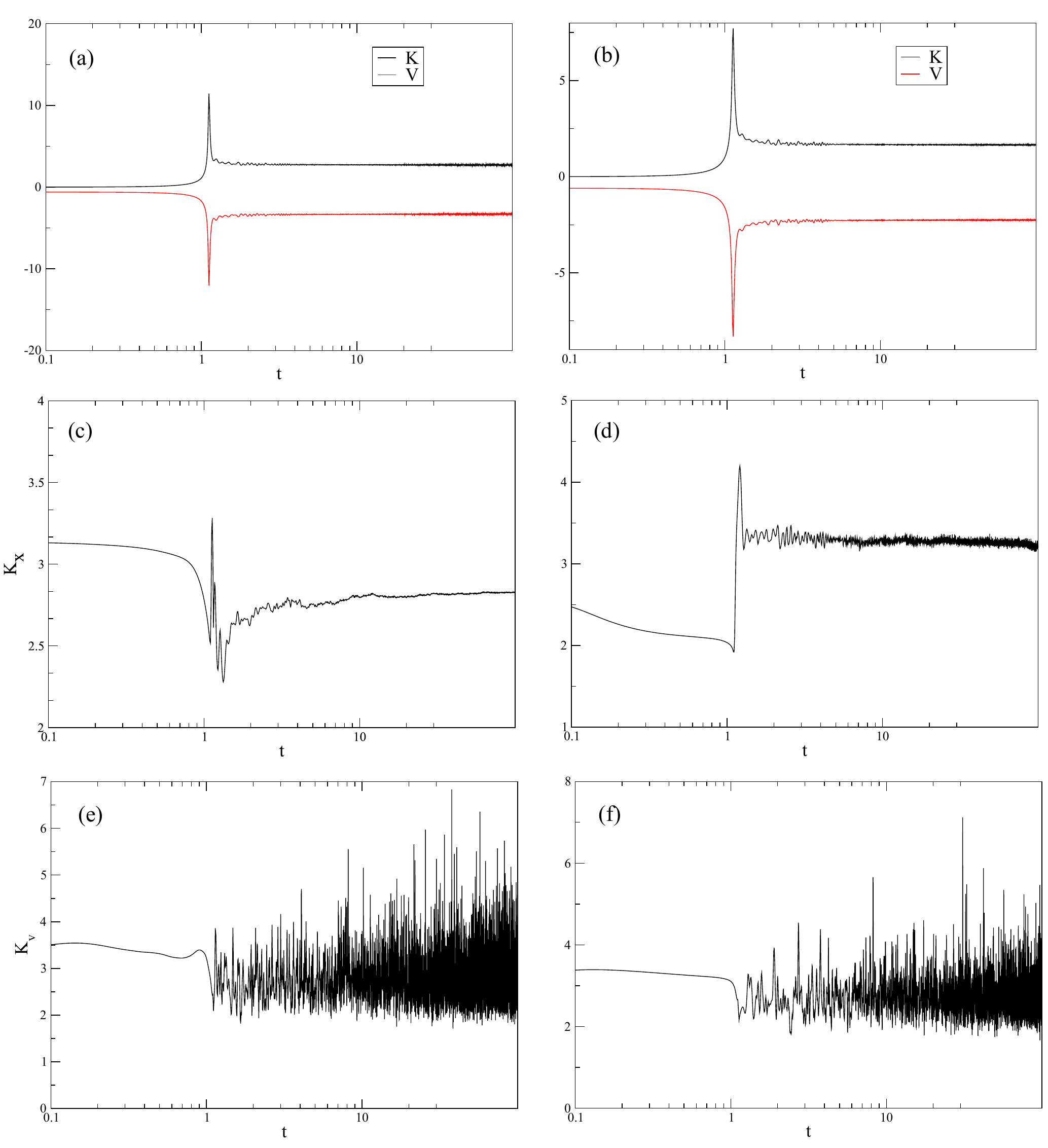}}}
\end{center}
\caption{Kinetic and potential energies per particle (Panel a) for the two-dimensional self-gravitating system with $N=16,384$ particles,
$R_1=1.0$, $R_2=1.5$ and $\epsilon=10^{-4}$.
Panel (c): Kurtosis of the $x$ component of the reduced position variable. Panel (e): Kurtosis of the $v_x$ component of the reduced velocity variable.
Panels (b), (d) and (f): Same as (a), (c) and (e) but with $\epsilon=10^{-3}$.}
\label{selfgravenerg}
\end{figure}

\begin{figure}[ptb]
\begin{center}
\scalebox{0.35}{\rotatebox{0}{\includegraphics{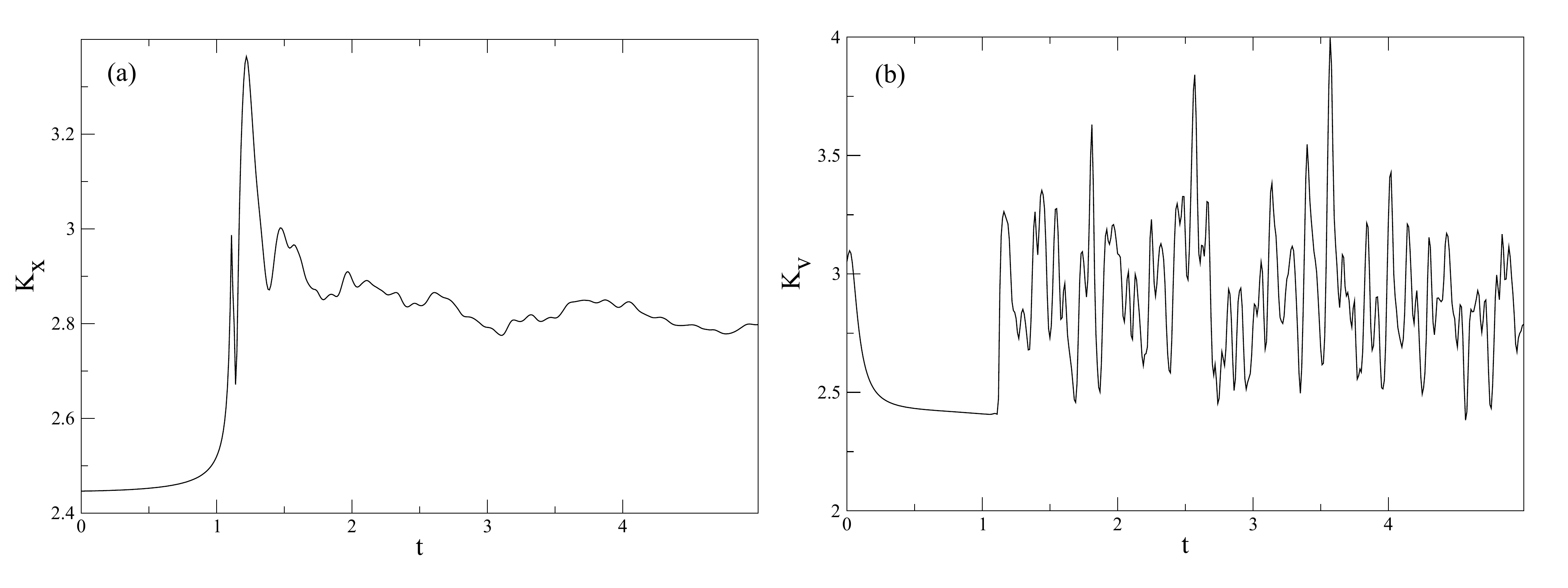}}}
\end{center}
\caption{Same as Fig.~\ref{selfgravenerg} but with $N=65,536$ and $\epsilon=10^{-3}$.}
\label{selfgravenerg_c}
\end{figure}

Similar results are obtained for the self-gravitating system.
Figure~\ref{selfgravenerg}a shows the time evolution of potential and kinetic energies with $N=16,384$, $R_1=1.0$, $R_2=1.5$ and $\epsilon=10^{-3}$, where the
so-called violent relaxation~\cite{lyndenbell} is clearly visible, with the final state close to the Gaussian equilibrium. Figures~\ref{selfgravenerg}b and~\ref{selfgravenerg}c
show the Kurtosis of the corresponding reduced variables in Eq.~(\ref{reducedvars}) for the $x$ components of the position ${\bf r}$ and velocity ${\bf v}$,
with $M=128$. The pattern observed is similar to that for the ring model. The position Kurtosis undergoes some change
during the contraction phase, and then settles down to an almost constant value, while the Kurtosis of velocity oscillates around the value $3$
corresponding the Gaussian distribution. It becomes apparent that there is no significant build up of correlations besides fluctuations, and
these maintain the same order of magnitude along all the time evolution. A similar behavior is observed in Figure~\ref{selfgravenerg} for $\epsilon=10^{-4}$.
The stronger deviations from the uncorrelated value occur around the maximum contraction during
the violent relaxation stage as expected due to higher values of the force for closer particles.
Figure~\ref{selfgravenerg_c} shows the results for a similar simulation with $N=65,536$ but with a shorter simulation time.
Their results are qualitatively the same but with smaller deviations from the corresponding Gaussian value.

\section{Quasi-stationary states and Vlasov instability}
\label{qssvi}

\begin{figure}[ptb]
\begin{center}
\scalebox{0.3}{\rotatebox{0}{\includegraphics{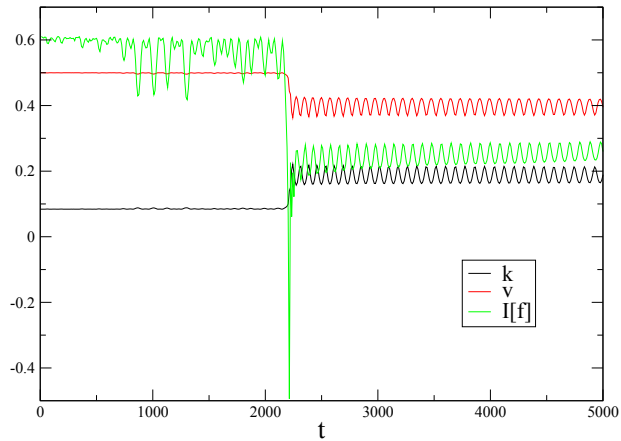}}}
\end{center}
\caption{Kinetic and potential energies and stability parameter $I[f]$ as defined in Eq.~\ref{stabcond}, for a homogeneous waterbag initial distribution,
$p_0=0.355$ and $N=10,000,000$.}
\label{stabillust}
\end{figure}

Quasi-Stationary States of the $N$-particle dynamics of long-range interacting systems
are now identified with stable stationary solutions of the Vlasov dynamics~\cite{proc1}
with a finite lifetime due to the loss of stability of the solution of the Vlasov
equation due to a slow secular time evolution of the one particle distribution function.
Since at any considered time the Vlasov equation gives an accurate
description of the dynamics up to terms of order $N^{-1}$ (provided all inter-particle correlations are dynamically generated)
it is valid to consider as initial condition for the Vlasov equation the distribution at any given time,
and therefore the stability of the slowly varying state is dictated by the Vlasov equation.
In Ref.~\cite{yamaguchi} it was shown for the HMF model that a homogeneous QSS with one-particle momentum distribution function $f(p)$
is stable if the following condition is satisfied:
\begin{equation}
I[f]\equiv1+\frac{1}{2}\int_{-\infty}^\infty\frac{f^\prime(p)}{p}\:dp>0.
\label{stabcond}
\end{equation}
From the considerations above, the lifetime of a homogeneous QSS is determined by the precise moment the secular evolution
of the distribution $f(p)$ due to cumulative effects of collisions (graininess) is such that $I[f]$ become negative, irrespective of the presence of any correlations
created from such collisions or how much time elapsed since the (uncorrelated) initial state. This is illustrated in a simulation with
$N=10,000,000$ particles in Figure~\ref{stabillust}.
The time value at which the QSS looses its stability is also precisely the moment
at which $I[f]$ in Eq.~(\ref{stabcond}) becomes negative.

\begin{figure}[ptb]
\begin{center}
\scalebox{0.35}{\rotatebox{-90}{\includegraphics{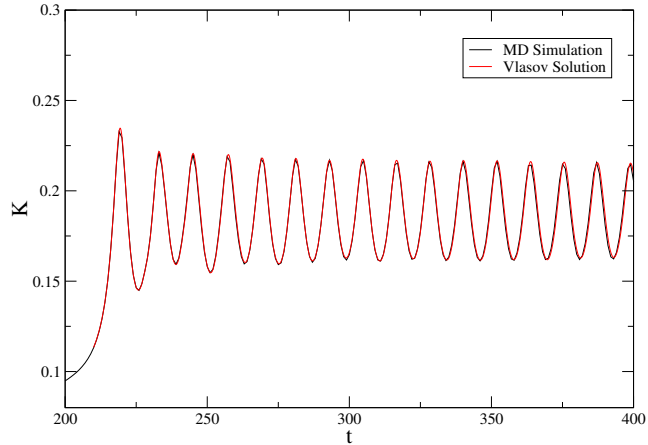}}}
\end{center}
\caption{Same as in fig.~\ref{stabillust} but $N=1,048,576$ and
the Vlasov simulation stating from the distribution function obtained from the final state of a MD simulation
at $t=210$. Both oscillate at almost the same frequency, with a slight shift as time evolves and are very close to one another.}
\label{stabcont}
\end{figure}

Even though the QSS has a finite lifetime due to small collisional terms,
the state after the loss of stability can still be used as an initial condition for the Vlasov equation
to describe the time evolution of the system for another finite time span.
As an example we consider a MD simulation with the same initial condition as in fig.~\ref{stabillust} but with $N=1,048,576$ particles and
stopped just after the loss of stability of the QSS.
Then the one-particle distribution function is determined from simulated data and used as the initial state for a Vlasov simulation.
The left panel in Figure~\ref{stabcont} shows the results for the Kinetic energy for both Vlasov and MD simulations starting from
this same initial state and both agree quite well.

\begin{figure}[ptb]
\begin{center}
\scalebox{0.6}{\rotatebox{0}{\includegraphics{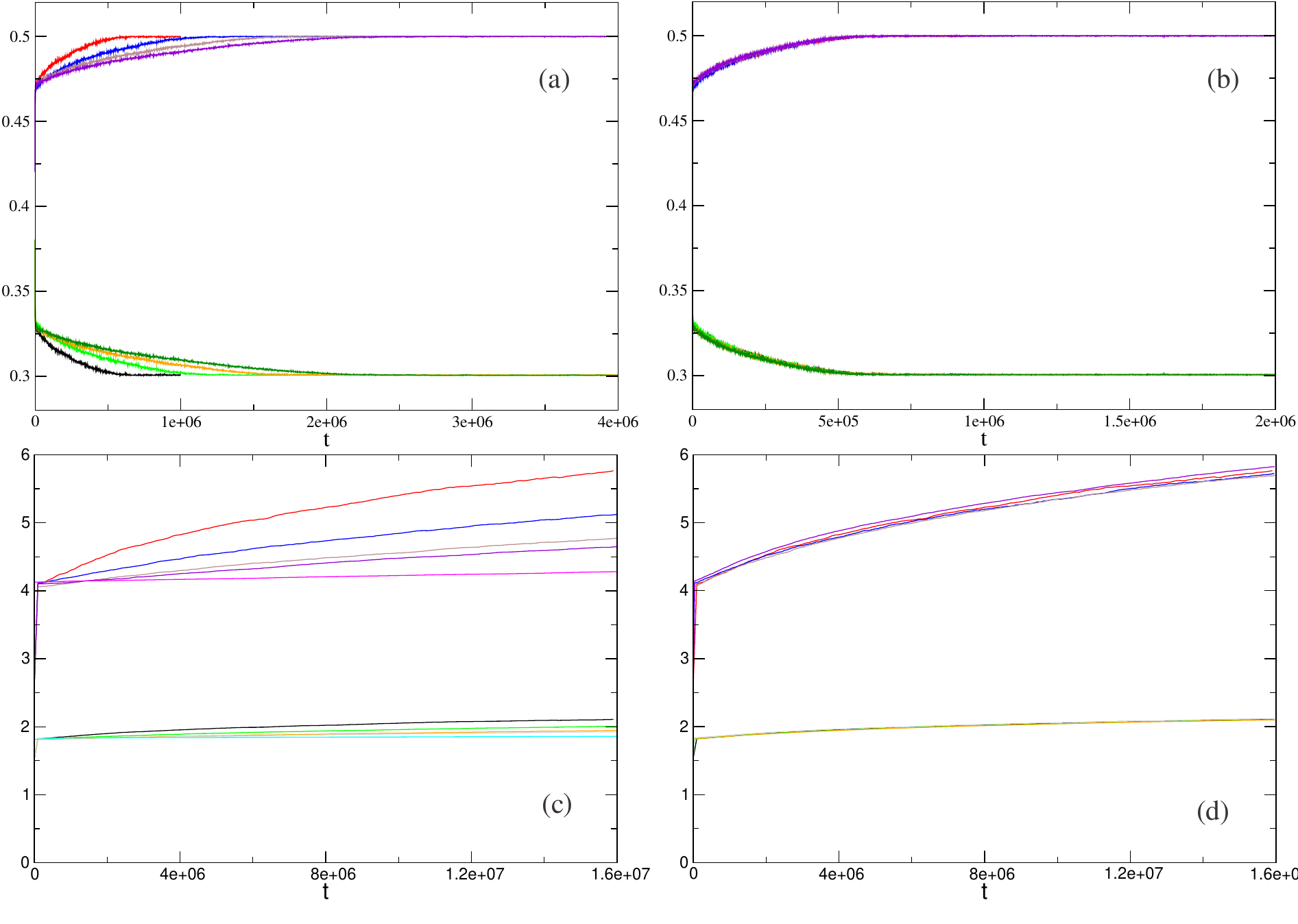}}}
\end{center}
\caption{Panel a) Evolution of kinetic $K$ (upper curves) and potential $V$ (lower curves) energies per particle for the three level
initial condition in Eq.~(\ref{threeini}) with
$p_1=0.3$, $p_2=1.8$, $f_1=0.0454$, $f_2=0.165$, $\theta_0=4.25$ corresponding for the values $N=100,000$, $N=200,000$, $N=300,000$ and $N=400,000$. The greater the value of $N$
the greater the time to attain the plateaus in the curves.
Panel b) Same as (a) but with a rescaling of time $t\rightarrow t/(N\times10^{-3})$ as expected for a non-homogeneous state.
Panel c) Fourth (lower curves) and sixth (upper curves) moments of the velocity distribution function for a three level initial condition with
$p_1=0.3$, $p_2=1.8$, $f_1=0.061$, $f_2=0.041$, $\theta_0=2\pi$ and for $N=40,000$, $N=60,000$, $N=80,000$, $N=100,000$ and $N=200,000$.
The greater $N$ the smaller the derivative of the moments at a given time.
Panel d) Same as (c) but with a time rescaling $t/(N/40,000)^2$. All curves collapse in a single curve for each moment.
}
\label{threeenerg}
\end{figure}

\begin{figure}[ptb]
\begin{center}
\scalebox{0.27}{{\includegraphics{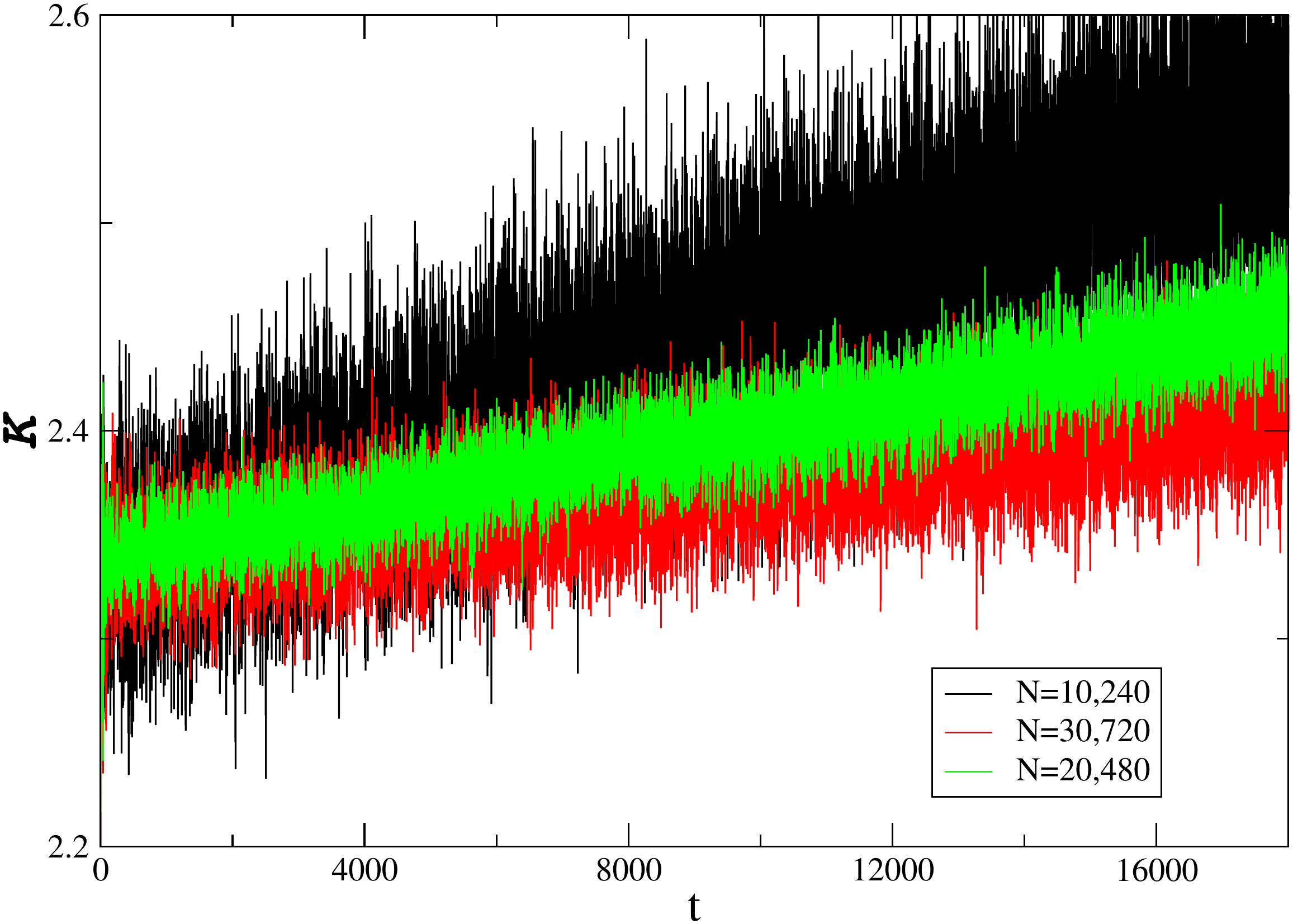}}}
\scalebox{0.27}{{\includegraphics{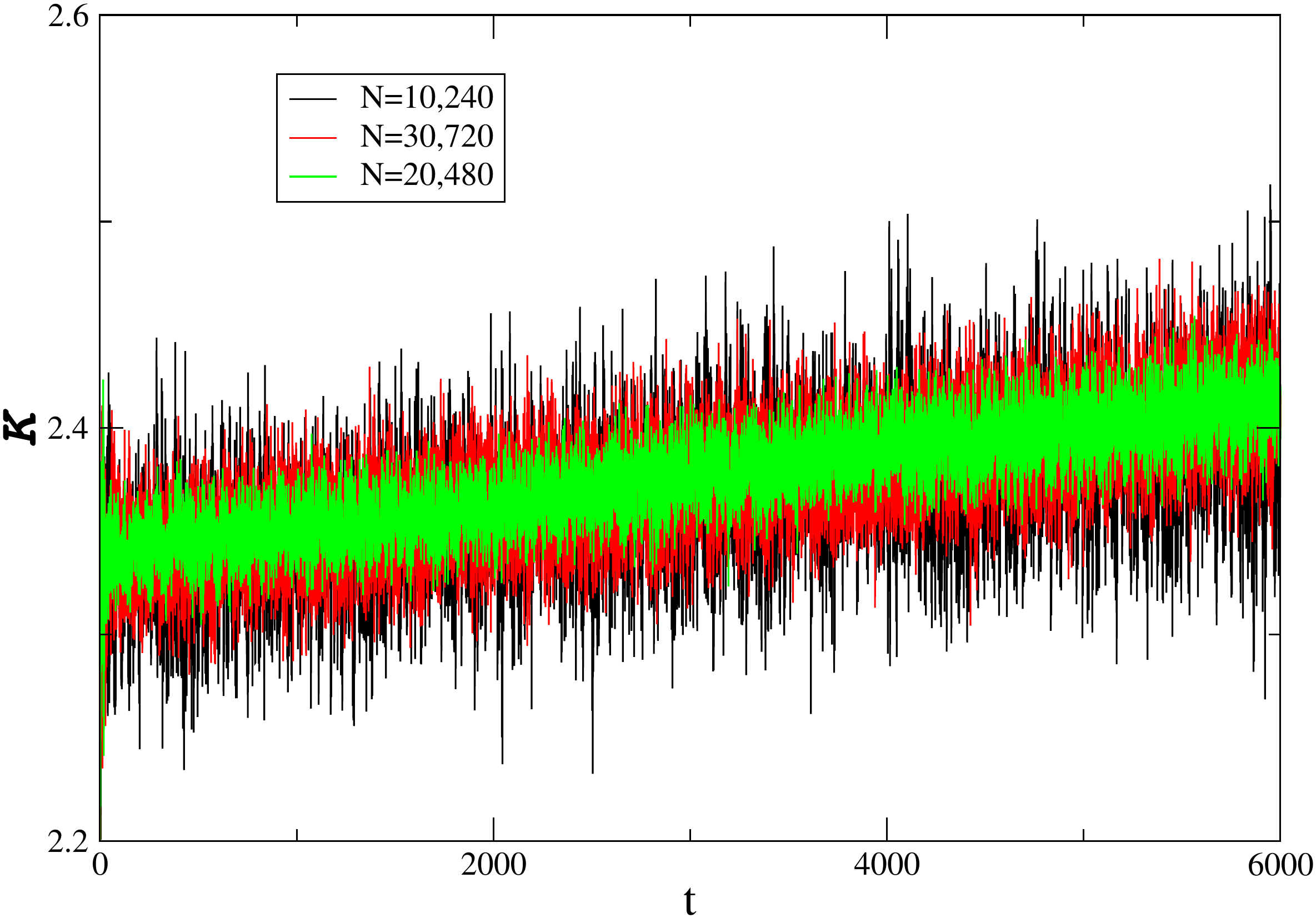}}}
\end{center}
\caption{Left Panel: Kurtosis as a function of time for the velocity distribution function of the 2-dimensional
self-gravitating system with a virialized initial condition and a spatially uniform distribution on a disc with
unit radius, for different number of particles. Right Panel: The same as the left panel but with a rescaling of time
$t\rightarrow t/(N/10240)$.}
\label{figfinal}
\end{figure}

For finite $N$ a QSS state is continually changing with time due to the
cumulative effects of collisions. A lifetime can be meaningfully ascribed only if a given property of the state changes abruptly,
as for instance in the case corresponding to Figure~\ref{stabillust} where the magnetization changes from a vanishing value associated with a homogeneous state
to a non-vanishing value. Nevertheless this abrupt change or dynamical phase-transition~\cite{nv3} is due to the loss of Vlasov stability
of the homogeneous QSS. This is not always the case and as an example of that let us consider the HMF model with a three level
distribution as an initial condition given by:
\begin{equation}
f_0(p,\theta)=\left\{
\begin{array}{ll}
f_1 & \mbox{if $|p|\leq p_1$ and $0<\theta<\theta_0$},\\
f_2 & \mbox{if $p_1<|p|<p_2$ and $0<\theta<\theta_0$},\\
0 & \mbox{otherwise},
\end{array}
\right.
\label{threeini}
\end{equation}
for $p_1$, $p_2$, $f_1$ and $f_2$ given constants. Panel (a) of Figure~\ref{threeenerg} shows the kinetic and potential energies per particle for a choice of
these constants. The state resulting from the violent relaxation is stable (this is shown if the figure where increasing values of $N$ lead to slower
variation of $K$ and $V$), but effects of collisions are nevertheless pronounced, is such a way that it is hardly possible to consider this state
as quasi-stationary. Panel (b) of the same figure shows that the collisional evolution after the violent relaxation scales with $N$ as expected
since in this case the kinetic equation is given by the Landau or Balescu-Lenard equations with a collisional integral proportional to $N$.
For homogeneous states the picture changes as can bee seen in Panel (c) of Figure~\ref{threeenerg}. Since the Kinetic and potential energies are almost constant
for the time interval considered, the slow secular evolution of the velocity distribution function is better grasped by considering its first moments
$M_k=\langle p^k\rangle$. The moments $M_4$ and $M_6$ are shown in Panel (d) with the cumulative effects of collisions clearly visible. As discussed in the
introduction the Landau and Balescu-Lenard collisional integrals vanish in the homogeneous states of one-dimensional systems, and consequently the scaling
of the time evolution with $N$ must be slower than for the inhomogeneous case. Panel (d) shows the same curves but with a rescaling proportional to $N^2$,
resulting in all curves for each momentum collapsing in a single curve. This as important result since it is at variance with the $N^{1.7}$ and $e^N$
scalings obtained in Refs.~\cite{nv1} and~\cite{nv2}, respectively, but is in agreement with the same scaling obtained for homogeneous one-dimensional
plasmas in Refs.~\cite{dawson} and~\cite{rouet}. For inhomogeneous states and higher dimensional systems this scaling is predicted from
well established physical theories, such as for the result shown in Fig.~\ref{figfinal} for a two-dimensional
self-gravitating system where a scaling proportional to $N$ is evident, in
accordance to the results in Ref.~\cite{marcos} but at variance to those in Ref.~\cite{levin0}.

There are some possibilities to explain those different scalings in the time evolution for the
homogeneous states of the HMF model. Either the scaling is state dependent since in our case we considered a three level initial state, while in
Refs.~\cite{nv1} and~\cite{rouet} a waterbag and a semi-elliptic initial distributions were used, respectively. This would imply that the $N$ dependence of
the collisional integral in the still unknown kinetic equation would vary with the statistical state, which in the authors opinion would be
a quite awkward case. Another possible explanation would be that since the author of Refs.~\cite{nv1} and~\cite{rouet} only considered small numbers
of particles in their simulations ($20,000$ at the most) and much smaller than the cases considered here, the $N^{1.7}$ and $e^N$ scaling would be
due to the small size of their systems. A separate publication will thoroughly explore such possibilities and discuss how to obtain
a kinetic equation valid for homogeneous states of one-dimensional systems.

\section{Concluding remarks}
\label{conclu}

In order to show how different interaction potentials lead to different convergence speed to the Vlasov (mean-field)
dynamics we have considered three different one-dimensional models with long-range interactions extensively studied
in the literature: the Hamiltonian Mean Field Model (HMF), the Self Gravitating Ring Model, and a 2-D systems of
Gravitating Particles (sheet model). All simulations for these models were performed starting from a waterbag
initial condition indicated in equations~\ref{wbinitheta} and~\ref{wbinix}. 

Simulations for the ring model are shown in figure~\ref{ringconv} for some values of the softening parameter $\epsilon$ and
convergence toward the Vlasov values of the kinetic energy gets clearly poorer for smaller values of this parameter, as the
interaction gets stronger at short distances where collisional effects are more important. For the self-gravitating
sheet model, an increasing agreement with increasing N is also obtained as given in figure~\ref{sheetconv}. The convergence to a
mean-field description is thus strongly affected by  the short range part of the force. The stronger the latter the
more important are the collisional effects, and the smaller the agreement in time window of the Vlasov equation
with the finite $N$ dynamics.

For sake of completeness and for the present paper to be self-contained, we discuss in the appendices the
derivation of the Vlasov equation from a re-summation of different contributions of a diagrammatic expansion.
It is shown that provided all  inter-particle correlations are dynamically created, they do not alter the
validity of the Vlasov equation. This means that deviations from the solution of the Vlasov equation are
not due to the building up of the correlations with time, since their order of magnitude is preserved by
the dynamics, but to the secular accumulation of small collisional corrections of order $1/N$ (see Section~\ref{secrole}).
The one-particle distribution function at any point of the evolution can thus be used as the initial condition to solve
the Vlasov equation, which is valid for a given time span starting at this value of time.

We also obtained a scaling as $N^2$ for the dynamics of a homogeneous state of the HMF model, at variance with previous results
that used smaller number of particles than in the present work~\cite{nv1,nv2}. For a two-dimensional system we obtained a scaling
in agreement with Ref.~\cite{marcos} but at variance with Ref.~\cite{levin0}.
This is a clear indication that a more comprehensive study if this problem us still lacking and will be a subject of another publication.

\section{Acknowledgments}

This work is partially financed by CNPq and CAPES (Brazilian government agencies).

\appendix
\section{Kinetic equation for large $N$ -- The Vlasov equation}
\label{appendixa}

Here we show that for long-range interacting system the Vlasov equation is kinetic equation describing the dynamics of the system
for a large number of particles $N$. The approach chosen here, i.~e.\ the dynamics of correlations is fully developed in
Balescu's monograph Ref.~\cite{balescu2} (see Ref.~\cite{liboff} for the notations used in this paper), which we succinctly present now. Its extension
for self-gravitating systems of non-identical particles is discussed in Appendix~\ref{appendixb}.

\subsection{Formal solution of the Liouville equation}
\label{apa1}

A generic Fourier coefficient can be decomposed as
\begin{equation}
a_{{\bf k}_{1},\ldots,{\bf k}_{l}}=a_{{\bf k}_1}a_{{\bf k}_2}\cdots a_{{\bf k}_l}+a_{[{\bf k}_1,\ldots,{\bf k}_l]},
\label{jan29.1}
\end{equation}
as a sum of a completely factored part and a pure correlation correlation represented as $a_{[{\bf k}_1,\ldots,{\bf k}_l]}$.
Using the completeness relation:
\begin{equation}
\int |\{{\bf k}\}\rangle\langle\{{\bf k}\}|\:d^N{\bf k}=\hat {\bf 1},
\label{completk}
\end{equation}
in Eq.~(\ref{rho}) we obtain:
\begin{eqnarray}
\lefteqn{a_{\{\bf k\}}({\bf v},t)=\langle\{{\bf k}\}|f_N(t)\rangle = -\frac{1}{2\pi i}\oint dz\:e^{-izt}
\left[R_{\bf k}(z)\:\delta^{Kr}(\{{\bf k}-{\bf k^\prime}\})\right.}\nonumber\\
 & & -R_{\bf k}(z)\:\langle\{{\bf k}\}|\delta\hat L|\{{\bf k^\prime}\}\rangle\:
R_{\bf k^\prime}(z)
\nonumber\\
 & & \left.+ \sum_{\{\bf k^{\prime\prime}\}}R_{\bf k}(z)\:\langle\{{\bf k}\}|\delta\hat L|\{{\bf k^{\prime\prime}}\}\rangle
\:R_{\bf k^{\prime\prime}}(z)\:\langle\{{\bf k^{\prime\prime}}\}|\delta\hat L|\{{\bf k^\prime}\}\rangle \:R_{\bf k^{\prime}}(z)\right.\nonumber\\
 & & +\cdots]\: a_{\{\bf k^\prime\}}({\bf v},0),
\label{braketexp}
\end{eqnarray}
with
\begin{equation}
R_ {\bf k}(z)=\frac{1}{i\left(\omega_{\bf k}-z\right)},\hspace{10mm}\omega_{\bf k}\equiv\sum_{j=1}^N{\bf k}_j\cdot{\bf v}_j,
\end{equation}
\begin{eqnarray}
 & & \langle\{{\bf k}\}|\delta\hat L|\{{\bf k^{\prime}}\}\rangle = \sum_{j<l}\delta L_{jl}(\{{\bf k}\},\{{\bf k}^\prime\}),\nonumber\\
 & & \delta L_{jl}(\{{\bf k}\},\{{\bf k}^\prime\}) =\nonumber\\
 & & \hspace{10mm}-\frac{i}{N}V_{|k_{j}-k_{j}^\prime|}(\mathbf{k}_{j}
-\mathbf{k}_{j}^{\prime })\:\partial_{jl}\:
\delta(\mathbf{k}_{j}+\mathbf{k}_{l}-\mathbf{k}_{j}^{\prime }
-\mathbf{k}_{l}^{\prime})\prod_{m\neq j,l}\delta ^{Kr}(\mathbf{k}_{m}-\mathbf{k}_{m}^{\prime }),
\nonumber\\
\label{dldef}
\end{eqnarray}
with $\partial_{ij}=\partial/\partial v_i-\partial/\partial v_j$, $\delta^{Kr}(\{{\bf k}\})=\prod_j \delta^{Kr}({\bf k}_j) $, $\delta^{Kr}(0)=1$, $\delta^{Kr}({\bf k}\neq0)=0$
and the Fourier transform of the potential given by:
\begin{equation}
V_{k}=\int V(r)e^{i\mathbf{k}\cdot \mathbf{r}}\:dr.
\label{fourierv}
\end{equation}
Note that $V_k$ depends only on $k=|{\bf k}|$ for a central potential.


\subsection{Diagrammatic Representation}
\label{apa2}

Each term in the expansion~(\ref{braketexp}) can be represented by a diagram according to the rules:
\begin{enumerate}[i]
\item For each element $R_{\bf k}(z)$ we associate a set of lines going from
right to left. The number of lines are the same as the number of non-vanishing
wave vectors in the set $\{\bf k\}$;
\item Each line has an index that represents the particle associated to the
wave-vector.
\item To each term $\delta L_{ij}(\{{\bf k}\},\{{\bf k}^\prime\})$ we
associate a vertex, with in and out lines concurrent with those with index $i$ and $n$
in the set $\{\mathbf{k}\}$ (if exists) and one of lines whit index $j$ and $n$
in the set $\{\mathbf{k}^\prime\}$ (if exists).
\item When considering the contribution to the reduced distribution $f_s$, for each vertex at least one of its two particles index
must belong to the set $\{1,\ldots,s\}$ or appear at a line at its left.
\end{enumerate}
Rule (iv) above comes from the presence of the operator $\partial_{ij}$ in the definition of $\delta L$ in Eq.~(\ref{dldef}),
leading to a vanishing surface term if this rule is not satisfied.

\begin{figure}[ht]
\begin{center}
\scalebox{0.8}{{\includegraphics{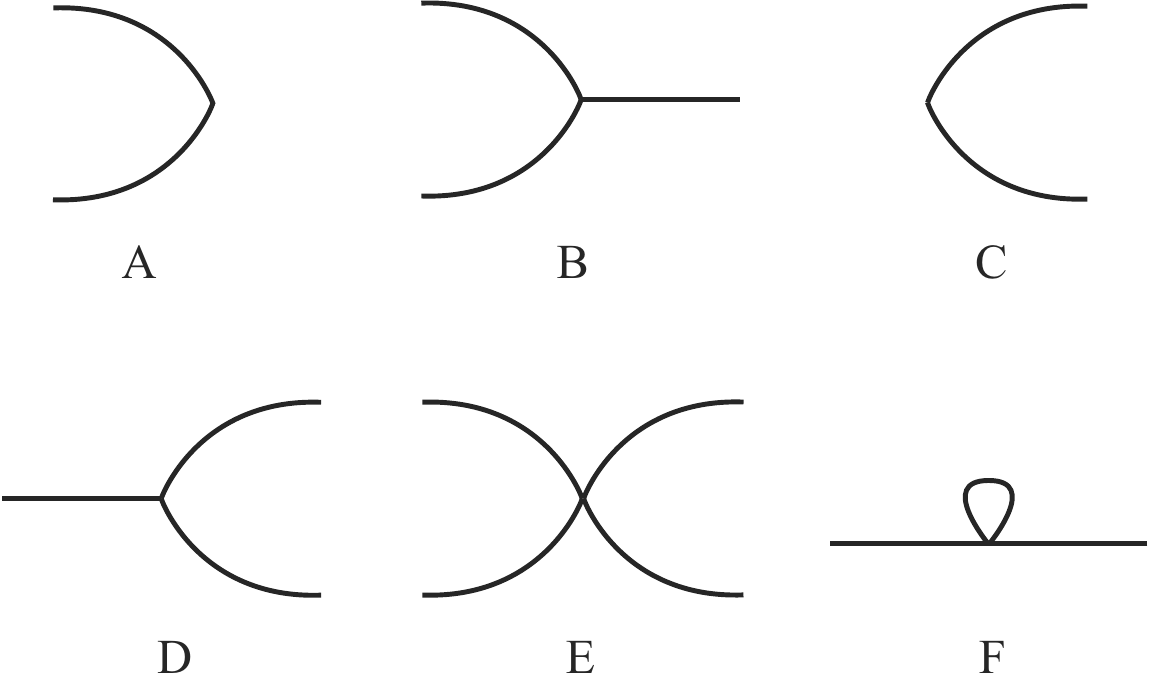}}}
\end{center}
\caption{Different vertices representing the interaction term in Eq.~(\ref{dldef}).}
\label{figA1}
\end{figure}

From Eq.~(\ref{dldef}) the wave-vectors change only two by two such that
${\bf k}_1+{\bf k}_2={\bf k}'_1+{\bf k}'_2$. In this way there are six possible vertices as shown in Fig.~\ref{figA1}.
To each line to the left and to the right of a vertex we associate a particle index.
Some examples of diagrams are given in Fig.~\ref{factex}.

In arguments of a Fourier coefficient we only write explicitly the velocities of those particles with a non-vanishing  wave-vector, and for the Fourier
coefficient of $s$-particle distribution functions a vanishing wave-vector for a given particle implies an integration over the coordinates (but not the velocities) of this particle,
we have for the Fourier coefficient of a two-particle distribution functions:
\begin{equation}
a_{\bf k}({\bf v}_\alpha;{\bf v}_j)=\int\:f_2({\bf r}_\alpha,{\bf v}_\alpha;{\bf r}_j,{\bf v}_j) e^{i{\bf k}\cdot{\bf r}_j} d{\bf r}_j=a_{\bf k}({\bf v}_\alpha)a_0({\bf v}_j).
\label{fourf2}
\end{equation}

\section{Systems with non-identical particles: Self-gravitating systems}
\label{appendixb}

For non-identical particles the $N$-particle distribution function $f_N$ is not symmetric, and the above approach cannot be
used as such. Let us consider now a subset of $s$-particles denoted by $\{i_{k}\}\equiv
\{i_{1},\ldots ,i_{s}\}$, with each $i_{k}$ different from the others
particle indices in the set and ranging from $1$ to $N$. The $s$-particle
distribution function for the set $\{i_{k}\}$ is defined by
\begin{equation}
f_{s}(\{i_{k}\})\equiv f_{s}(i_{1},\dots ,i_{s})=\int f_{N}(1,\ldots ,N)\:dj_{1}\cdots dj_{N-s},
\label{reducedf}
\end{equation}%
where $\{j_{l}\}$ is the set of particle indices from the $N$-particle set
not contained in $\{i_{k}\}$. By integrating both sides of Eq.~(\ref%
{liouvilleeq}) over the coordinates and momenta of particles in $\{j_l\}$,
and discarding surface terms, we obtain:
\begin{equation}
\left[ \frac{\partial }{\partial t}-\hat L_{s}\right] F_{s}(\{i_{k}\})=%
\sum_{k=1}^{s}\sum_{l=1}^{N-s}\frac{\partial }{\partial \mathbf{p}_{i_{k}}}%
\cdot \int \frac{\partial V_{i_{k}j_{l}}}{\partial \mathbf{r}_{i_{k}}}%
F_{s+1}(\{i_{k}\},j_{l})\:dj_{l},
\label{bbgky}
\end{equation}%
where $\hat L_{s}$ is the Liouville operator  for the $s$-particle subsystem and $F_{s+1}(\{i_{k}\},j_{l})$ stands for
$F_{s+1}(i_{1},\ldots,i_{s},j_{l})$. At this point it is important to notice that since we are
considering the possibility of a system with different masses, the reduced
functions as defined by Eq.~(\ref{reducedf}) are not symmetric by
permutation of two particles and Eq.~(\ref{bbgky}) is the final form of the
BBGKY hierarchy.

For a self-gravitating system of particles with different masses, the
potential energy is written as
\begin{equation}
V_{ij}=V(|\mathbf{r}_{i}-\mathbf{r}_{j}|)=-Gm_{i}m_{j}\frac{\mathbf{r}_{i}-
\mathbf{r}_{j}}{|\mathbf{r}_{i}-\mathbf{r}_{j}|^{3}}\equiv m_{i}m_{j} h({\bf r}_i-{\bf r}_j),
\label{gravpotenerg}
\end{equation}
where $h({\bf r}_i-{\bf r}_j)$ is the force between two particles of unit mass at positions ${\bf r}_i$ and ${\bf r}_j$.
This particular form can be used to further simplify the hierarchy in Eq.~(\ref{bbgky}).
To illustrate how to proceed we consider the case $s=1$. We
first define the one-particle mass-density in phase space:
\begin{equation}
f_{1}(\mathbf{r},\mathbf{p})=\sum_{i=1}^{N}m_{i}F_{1}^{(i)}(\mathbf{r},
\mathbf{p}),  \label{massdensity}
\end{equation}
where $F_{1}^{(i)}(\mathbf{r},\mathbf{p})$ is given by $F_{1}(\mathbf{r}_{i},
\mathbf{p}_{i})$ computed at $\mathbf{r}_{i}=\mathbf{r}$ and $\mathbf{p}_{i}=\mathbf{p}$. Similarly we define
\begin{equation}
f_{2}(\mathbf{r},\mathbf{p},\mathbf{r}^{\prime },\mathbf{p}^{\prime})
=\left. \sum_{i\neq j=1}^{N}m_{i}m_{j}F_{2}(i,j)\right\vert _{i=\mathbf{r},\mathbf{p},:j=\mathbf{r}^{\prime },\mathbf{p}^{\prime }},
\label{twomassdens}
\end{equation}%
which as defined is symmetric by permutation of $\mathbf{r},\mathbf{p}$ and $\mathbf{r}^{\prime },\mathbf{p}^{\prime}$.
Using Eq.~(\ref{bbgky}) for $s=1$ and eqs.~(\ref{massdensity}) and~(\ref{twomassdens}), it is straightforward to show that
\begin{equation}
\left[ \frac{\partial }{\partial t}+\mathbf{v}\cdot \frac{\partial }{\partial \mathbf{r}}\right] f_{1}(\mathbf{r},\mathbf{p})=
-\frac{\partial }{\partial \mathbf{v}}\cdot \int \mathbf{h}(\mathbf{r}-\mathbf{r}^{\prime})
f_{2}(\mathbf{r},\mathbf{p},\mathbf{r}^{\prime },\mathbf{p}^{\prime })\:d\mathbf{r}\:d\mathbf{p}.
\label{eqforrho1}
\end{equation}
A similar simplification is not possible for $s\geq3$ due to the operator $\hat L_s$ in the left-hand side of Eq.~(\ref{bbgky}).
From the discussion above on identical particles, it is reasonable to suppose that correlations between particles are of order $N^{-1}$
and therefore negligible for large $N$. For a system of self-gravitating identical particles this statement can be proved using
the diagrammatic method in the same lines as for a plasma~\cite{balescu2}.

\end{document}